\documentclass{aa}
\usepackage{graphicx}
\usepackage{txfonts}

\newcommand{\MSTAR}{\mbox{$M_{\star}$}}
\newcommand{\LSTAR}{\mbox{$L_{\star}$}}
\newcommand{\TEFF}{\mbox{$T_{\rm eff}$}}

\newcommand{\MSOL}{\mbox{$M_{\sun}$}}
\newcommand{\LSOL}{\mbox{$L_{\sun}$}}

\newcommand{\micron}{\mbox{$\mu$m}}
\newcommand{\KMS}{\mbox{km s$^{-1}$}}
\newcommand{\HOH}{\mbox{H$_2$O}}
\newcommand{\PERSQCM}{\mbox{cm$^{-2}$}}

\newcommand{\VMICRO}{\mbox{$\varv_{\rm micro}$}}

\newcommand{\LOGG}{\mbox{$\log \varg$}}
\newcommand{\ABUNDSI}{\mbox{$\log A_{\rm Si}$}}

\newcommand{\ABUNDH}{\mbox{$\log A_{\rm H}$}}
\newcommand{\SIISOTOPEF}{\mbox{$^{28}$Si/$^{29}$Si}}
\newcommand{\SIISOTOPES}{\mbox{$^{28}$Si/$^{30}$Si}}
\newcommand{\alfboo}{\mbox{$\alpha$~Boo}}
\newcommand{\gamcru}{\mbox{$\gamma$~Cru}}
\newcommand{\siglib}{\mbox{$\sigma$~Lib}}
\newcommand{\vcen}{\mbox{V806~Cen}}
\newcommand{\epsmus}{\mbox{$\varepsilon$~Mus}}
\newcommand{\tetaps}{\mbox{$\theta$~Aps}}
\newcommand{\alfori}{\mbox{$\alpha$~Ori}}
\newcommand{\alfsco}{\mbox{$\alpha$~Sco}}
\newcommand{\alfher}{\mbox{$\alpha$~Her}}
\newcommand{\vycma}{\mbox{VY~CMa}}
\newcommand{\vxsgr}{\mbox{VX~Sgr}}
\newcommand{\raql}{\mbox{R~Aql}}
\newcommand{\rhya}{\mbox{R~Hya}}
\newcommand{\whya}{\mbox{W~Hya}}
\newcommand{\lpup}{\mbox{L$_2$~Pup}}
\newcommand{\gcirs}{\mbox{GCIRS3}}
\begin{document}
\title{
High spectral resolution spectroscopy of the SiO fundamental lines 
in red giants and red supergiants with VLT/VISIR
\thanks{
Based on VISIR observations made with the Very Large Telescope of 
the European Southern Observatory. Program ID: 087.D-0522(A)}
}

\author{K.~Ohnaka
}

\offprints{K.~Ohnaka}

\institute{
Max-Planck-Institut f\"{u}r Radioastronomie, 
Auf dem H\"{u}gel 69, 53121 Bonn, Germany\\
\email{kohnaka@mpifr.de}
}

\date{Received / Accepted }

\abstract
{
The mass-loss mechanism in red giants and red supergiants is not yet 
understood well.  
The SiO fundamental lines near 8~\micron\ are potentially useful for probing 
the outer atmosphere, which is essential for clarifying 
the mass-loss mechanism.  However, these lines have been little explored 
till now. 
}
{
We present high spectral resolution spectroscopic observations of 
the SiO fundamental lines near 8.1~\micron\ in 16 bright red giants and red 
supergiants.  Our sample consists of seven normal (= non-Mira) K--M giants 
(from K1.5 to M6.5), three Mira stars, three optically 
bright red supergiants, two dusty red supergiants, and 
the enigmatic object \gcirs\ near the Galactic center.  
}
{
Our program stars were observed between 8.088 and 8.112 \micron\ 
with a spectral resolution of 30\,000 using VLT/VISIR.   
}
{
We detected SiO fundamental lines in all of our program stars 
except for \gcirs.  
The SiO lines in normal K and M giants as well as optically bright 
(i.e., not dusty) red supergiants do not 
show P-Cyg profiles or blueshifts, 
which means the absence of systematic outflows 
in the SiO line forming region.  
On the other hand, we detected P-Cyg profiles in the SiO lines in the dusty red 
supergiants \vycma\ and \vxsgr, with the latter being a new detection.  
These SiO lines originate in the outflowing gas with the thermal dust 
continuum emission seen as the background.  
The outflow velocities of the SiO line forming region in \vycma\ and \vxsgr\ 
are estimated to be 27 and 17~\KMS, respectively.  
We derived basic stellar parameters (effective temperature, surface 
gravity, luminosity, and mass) for the normal K--M giants and optically bright 
red supergiants in our sample and compared the observed VISIR spectra with 
synthetic spectra predicted from MARCS photospheric models.  
Most of the SiO lines observed in the program stars warmer than $\sim$3400~K 
are reasonably reproduced by the MARCS models, which allowed us to 
estimate the silicon abundance as well as the \SIISOTOPEF\ and 
\SIISOTOPES\ ratios. However, we detected possible absorption excess in 
some SiO lines. 
Moreover, 
the SiO lines in the cooler red giants and red supergiant cannot be explained 
by the MARCS models at all even if the dust emission is taken into account. 
This disagreement may be a signature of the dense, extended molecular outer 
atmosphere.  
}
{}

\keywords{
infrared: stars --
techniques: spectroscopic -- 
stars: AGB, post-AGB -- 
stars: late-type -- 
stars: atmospheres -- 
stars: fundamental parameters
}   

\titlerunning{
High-spectral resolution spectroscopy of the SiO fundamental lines 
in red giants and red supergiants  
}
\authorrunning{Ohnaka}
\maketitle

\begin{table*}
\begin{center}
\caption {Summary of the VISIR observations.  
Seeing is in the visible. Int. time = Integration time. 
References for the heliocentric systemic velocities: 
$a$: Gontcharov (\cite{gontcharov06}), 
$b$: Hatzes \& Cochran (\cite{hatzes93}), 
$c$: Murdoch et al. (\cite{murdoch92}), 
$d$: Kerschbaum \& Olofsson (\cite{kerschbaum99}), 
$e$: O'Gorman et al. (\cite{ogorman12}), 
$f$: Smith et al. (\cite{smith89}), 
$g$: Kemper et al. (\cite{kemper03}), 
$h$: Young (\cite{young95}). 
}
\vspace*{-2mm}

\begin{tabular}{r l l r c c c c r c l}\hline
\# & Name  & Sp. Type & D.E.C. & $V_{\rm helio}$ & $t_{\rm obs}$ & Seeing   &
Airmass & Int. time & 
Calibrator & Remarks\\ 
   &       &        &              &  (\KMS) & (UTC)        & (\arcsec)&         & (sec)            &
           & \\
\hline

\hline
\multicolumn{11}{l}{K--M giants}\\
\hline

1 & \alfboo & K1.5III & 0.0  & $-5.2\pm0.3^{a,b}$ & 03:44:15 & 0.72 & 1.49 & 180 &
C8 & \\ 
2 & \gamcru & M3.5III & 0.0 & $21.0\pm0.6^{a,c}$ & 01:53:35 & 1.15 & 1.24 & 180 &
C4& \\ 
3 & \siglib & M3/M4III & 0.0 & $-3.9\pm0.6^a$ & 06:24:15 & 0.71 & 1.01 & 180 &
C6 & \\ 
4 & \vcen & M4.5III & 0.01  & $40.7\pm0.6^a$ & 02:08:55 & 1.39 & 1.20 & 180 & C4
& \\ 
5& \epsmus & M5III & $-0.01$ & $7.1\pm0.6^a$ & 04:28:15 & 0.90 & 1.42 & 180 &
C5&\\ 
6& \lpup & M5III & 0.60 & $52.1\pm0.3^d$ & 00:16:55 & 0.88 & 1.20 & 180 & C1 &
\\ 
7& \tetaps & M6.5III & 0.29 & $9.0\pm0.3^d$ & 02:56:55 & 1.16 & 1.70 & 600 &
C4&\\ 
\hline
\multicolumn{11}{l}{Red supergiants}\\
\hline
8 & \alfori & M2Iab & 0.25 & $19.6\pm0.2^e$ & 00:00:55 & 0.97 & 1.86 & 180 & C3
&\\ 
9 & \alfsco & M1.5Iab-b & 0.06 & $-3.5\pm5^{a,f}$ & 05:19:35 & 0.91 & 1.12 & 180 &
C6 & \\ 
10 & \alfher & M5Ib-II & $-0.04$ & $-30.4\pm5^{a,f}$ & 07:36:15 & 0.59 & 1.30 &
180 & C8 & \\ 
11& \vycma & M2.5-5Iae & 25 & $45.5\pm4.5^g$ & 01:10:15 & 0.98 & 1.31 & 180 &
C2&\\ 
12& \vxsgr & M4Ia-10eIa & 0.73 & $-5.6\pm1.3^g$ & 06:39:35 & 0.84 & 1.17 & 180 &
C6 &\\ 
\hline
\multicolumn{11}{l}{Mira stars}\\
\hline
13& \raql & M5-9e & 0.36  & $28.1\pm0.2^h$ & 07:50:15 & 0.63 & 1.40 & 180 & C8 & 
phase = 0.18\\ 
14& \rhya & M6-9eS & 0.24  & $-11.9\pm0.2^h$ & 05:49:35 & 0.83 & 1.07 & 180 & C6
& phase = 0.44\\ 
15& \whya & M7.5-9ep & 0.10 & $37.9\pm0.2^h$ & 02:41:35 & 1.34 & 1.11 & 180 & C5 & 
phase = 0.14\\ 
\hline
\multicolumn{11}{l}{Unknown nature}\\
\hline
16 & \gcirs &   & ---   & --- & 07:10:55 & 0.66 & 1.05 & 600 & C7 &
\\
\hline
\multicolumn{11}{l}{Calibrators}\\
\hline
C1 & $\alpha$ Car & F0II & --- & --- & 23:44:15 & 0.99 & 1.30 & 180 &  & \\
C2 & $\alpha$ CMi & F5IV-V & ---&---&  00:32:15 & 0.85 & 1.37 & 540 &  & \\
C3 & $\alpha$ CMi & F5IV-V & ---&---&  01:26:55 & 1.09 & 1.69 & 540 &  & \\
C4 & $\alpha$ Cen A & G2V & ---&---& 02:27:35 & 0.68 & 1.48 & 180 &  & \\
C5 & $\alpha$ Cen A & G2V & ---&---& 03:20:55 & 0.73 & 1.35 & 180 &  & \\
C6 & $\alpha$ Cen A & G2V & ---&---& 05:34:15 & 0.75 & 1.24 & 180 &  & \\
C7 & $\alpha$ Cen A & G2V & ---&---& 06:54:15 & 0.71 & 1.29 & 180 &  & \\
C8 & $\alpha$ Aql & A7V & --- & --- & 08:04:55 & 0.66 & 1.48 & 1400 &  & \\
\hline

\label{obslog}
\vspace*{-7mm}

\end{tabular}
\end{center}
\end{table*}

\section{Introduction}
\label{sect_intro}

The mass loss in red giants and red supergiants 
is a long-standing problem in stellar astrophysics.
The combination of the large-amplitude stellar pulsation and the momentum 
transfer from radiation to dust is often believed to be responsible 
for the mass loss (see, e.g., H\"ofner \cite{hoefner11} for recent 
review).  
However, as H\"ofner (\cite{hoefner11}) notes, there are stars with 
stellar parameters outside the range for this ``pulsation-enhanced 
dust-driven wind scenario'', for example, stars with much smaller pulsation 
amplitudes and/or with no or little dust.   
This suggests that we may not yet understand major physical processes
responsible for driving mass outflows in red giants and red supergiants 
in general.

For understanding the mass-loss mechanism, it is essential to study the 
region between the photosphere and the innermost region of the circumstellar
envelope, where the wind acceleration takes place. 
SiO is one of the important molecules in the atmosphere and circumstellar 
envelope of oxygen-rich cool evolved stars.  
Hinkle et al. (\cite{hinkle76}) revealed significant time variations in 
the SiO first overtone bands near 4~\micron\ in Mira stars.  
The SiO first overtone bands have also been used for testing 
the model atmospheres of red giants (e.g., Aringer et al. \cite{aringer99}; 
Lebzelter et al. \cite{lebzelter01}).  
Tsuji et al. (\cite{tsuji94}) determined the elemental abundance of silicon as 
well as its isotope ratios in M giants and M supergiants from 
high resolution spectra of the SiO first overtone lines.  
Their analysis also reveals that while the weak and moderate SiO lines 
can be explained well 
by the hydrostatic photospheric models, the strong SiO lines 
in the latest M giants (M7--8III) show excess emission 
originating in an extended outer atmosphere. 
This extended outer atmosphere---the so-called MOLsphere 
(Tsuji \cite{tsuji00b})---cannot be accounted for by hydrostatic 
photospheric models.  The presence of the MOLsphere in normal (= non-Mira) 
K--M giants, Mira stars, as well as 
red supergiants has been confirmed by IR spectroscopic and interferometric 
observations of molecules such as CO and \HOH\ (e.g., Tsuji et al. 
\cite{tsuji97}; Tsuji \cite{tsuji00a}, \cite{tsuji00b}, \cite{tsuji01}; 
Perrin et al. \cite{perrin04a}, \cite{perrin04b}, \cite{perrin05}; 
Wittkowski et al. \cite{wittkowski07}, \cite{wittkowski08}, 
\cite{wittkowski11}; 
Woodruff et al. \cite{woodruff08}, \cite{woodruff09}; 
Le Bouquin et al. \cite{lebouquin09}; 
Mart\'i-Vidal \cite{marti-vidal11}; 
Ohnaka \cite{ohnaka04a}, \cite{ohnaka04b}, \cite{ohnaka12}, 
\cite{ohnaka13}; 
Ohnaka et al. \cite{ohnaka11}, \cite{ohnaka13alfsco}).

The current MOLsphere models consist of ``ad hoc'' layers placed above the 
photosphere, and the density, temperature, and radius of the 
extended molecular outer atmosphere are determined to fit observed data.  
On the other hand, 
dynamical model atmospheres have been developed for Mira stars 
(e.g., Ireland et al. \cite{ireland11}; Bladh et al. \cite{bladh13}), 
which predict the density and temperature distribution, as well as 
molecular abundances in the outer atmosphere in a self-consistent manner.  
Such models can be confronted with observed spectra.  
Unfortunately, dynamical model atmospheres for normal 
red giants and red supergiants, whose variability amplitudes are much 
smaller than in Mira stars, have been less studied.  
Probing the physical properties of the outer atmosphere using the ad hoc 
models is useful for gleaning information about the as-yet-unidentified 
physical process responsible for forming the extended molecular outer 
atmosphere.

The SiO fundamental lines near 8 \micron\ are one of the ideal probes 
to study the physical properties of the outer atmosphere and the innermost 
circumstellar envelope.  
Geballe et al. (\cite{geballe79}) observed the SiO fundamental lines 
in the dusty red supergiant \vycma\ with a spectral resolution of 13000.
They discovered SiO lines with the P-Cyg profile originating in an 
expanding circumstellar envelope.  
The SiO lines with the P-Cyg profile in \vycma\ were also detected much later 
by Richter et al. (\cite{richter01}) with a higher spectral resolution of 
100\,000.  
Keady \& Ridgway (\cite{keady93}) detected
weak, blueshifted SiO lines in the dusty carbon star IRC+10216 originating
in the inner envelope.
These studies suggest that the SiO fundamental lines are potentially useful for
probing the outer atmosphere and the innermost circumstellar envelope.  
However, there are only few high resolution spectroscopic studies
of the SiO fundamental lines in red giants and red supergiants in the 
literature.
In this paper, we present the results of high resolution spectroscopic 
observations of the poorly explored SiO fundamental lines in a sample 
of red giants and red supergiants with VLT/VISIR.

\section{Observations}
\label{sect_obs}

\subsection{Sample}

Our sample consists of seven normal (= non-Mira-type) K--M giants, 
three Mira stars, five red supergiants, 
and the enigmatic object toward the Galactic center \gcirs. 
We selected these stars to cover a range of the spectral type, 
luminosity class, and variability, in order to 
systematically study the dependence of the SiO lines on the stellar 
parameters. 
As summarized in Table~\ref{obslog}, 
seven K--M giants in our sample cover the spectral type from K1.5III 
to M6.5III.  All of these stars except for \lpup\ and 
\tetaps\ show no or very little 
dust, as shown by the dust emission contrast (D.E.C.), which is 
defined as the ratio of the dust emission to the emission from the star 
integrated over the spectral range between 7.67 and 14.03~\micron\   
(Sloan \& Price \cite{sloan98}).  
The D.E.C values in Table~\ref{obslog} were taken from 
Sloan \& Price (\cite{sloan98}) except for \alfboo, \gamcru, and \vycma, 
which are not included in their list.  
We estimated the D.E.C. values of \alfboo\ and \gamcru\ to be zero 
based on the absence of an IR excess up to $\sim$1000~\micron\ 
(Dehaes et al. \cite{dehaes11}).  
We estimated the D.E.C value of \vycma\ from 
the model fitting of the spectral energy distribution presented 
by Harwit et al. (\cite{harwit01}).  
The negative D.E.C. values for \epsmus\ and \alfher\ from Sloan \& Price 
(\cite{sloan98}) probably result from the uncertainty in defining the 
stellar continuum.  We interpret them as showing no dust emission.  
Our sample includes three Mira stars, \rhya, \raql, and 
\whya.  While the last star is classified as a semiregular 
variable on Simbad, it is often regarded as Mira-like because 
its variability amplitude is larger than usual semiregular variables, 
and it shows clear periodicity (e.g., Woodruff et al. \cite{woodruff09}; 
Zhao-Geisler et al. \cite{zhao-geisler11}).  
The variability phase at the time of our VISIR observations was 
estimated from the light curves available at the American Association of 
Variable Star Observers (AAVSO).  
The red supergiants in our sample consist of three optically bright 
stars (\alfori, \alfsco, and \alfher) 
and two dusty stars (\vycma\  and \vxsgr). 
The nature of the last object in our sample, \gcirs, is not clear yet. 
Based on high spatial resolution mid-IR observations, 
Pott et al. (\cite{pott08}) suggest that it may be a dust-enshrouded carbon 
star, despite the strong silicate absorption at 10~\micron.  
We included this object, because some molecular lines may reveal 
the nature of the object.

\subsection{VLT/VISIR observations and data reduction}

We observed our program stars on 2011 April 19 between 8.088 and 
8.112~\micron\ with VLT/VISIR (Lagage et al. \cite{lagage04}). 
We used the high resolution long-slit (HR) mode with a slit 
width of 0\farcs4, which resulted in a spectral resolution 30\,000.    
We selected this spectral window, because it is not severely affected 
by the telluric lines, except for the strong ones at 8.091 and 8.107~\micron. 
A summary of the observations of the science targets and the calibrators 
is given in Table~\ref{obslog}.  
The data were reduced with the VISIR pipeline ver3.0.0 provided by ESO 
(Lundin et al. \cite{lundin08}).  The output of the pipeline 
reduction is the 1-D spectra of the targets.  
The wavelength calibration was carried out using the telluric lines.  
The uncertainty in the wavelength calibration is 
$\sim \! 7.0 \times 10^{-5}$~\micron\ (2.6~\KMS).  
The telluric lines were removed by dividing the spectra of the 
science targets with those of the calibrators.  In most cases, we used 
the calibrators that were observed before or after the science target 
within 0.5--1~hour at similar airmasses (difference smaller than $\sim$0.2) .  
In the case of \alfboo, 
it was necessary to use the calibrator observed 
with a time interval as long as 4.5~hours.  
In either case, however, the strong telluric lines at 8.091 and 
8.107~\micron\ could not be removed well, and therefore, we excluded these 
wavelengths from the analysis.  
The spectra are normalized by the local continuum level defined by the 
highest flux points in the observed spectral window.  
Because most of our program stars are very bright, we achieved S/N ratios of 
100--200.  The exception is \gcirs, on which we achieved an S/N ratio 
of only 20.

We converted the wavelength scale to the heliocentric frame 
and then to the laboratory frame.  
The heliocentric systemic velocities of our program stars are 
listed in Table~\ref{obslog} with the uncertainties.  
We took the systemic velocities measured by radio CO observations in 
the literature if available (references are also listed in the table).  
The systemic velocities derived from the CO lines originating in the 
expanding circumstellar envelope are not subject to atmospheric motions 
induced by pulsation and/or convection, which can affect the determination 
of the systemic velocity using spectral lines in the visible.  
In case velocities derived from different radio CO lines are 
available, we took the mean of the velocities and adopted a half of the 
maximum and minimum velocities as the uncertainty.  
For \alfboo, \gamcru, \siglib, \vcen, \epsmus, \alfsco, and \alfher, 
no radio CO observations are available.  
For these stars, 
we took the radial velocities from Gontcharov (\cite{gontcharov06}), 
which were derived from spectral lines in the visible. 
These data represent the systemic velocity well for K and early/mid-M giants 
whose variability amplitude is small.  
For example, the time variation in the radial velocity in \alfboo\ and \gamcru\ 
is $\pm0.3$ and $\pm0.6$~\KMS, respectively (Hatzes \& Cochran 
\cite{hatzes93}; Murdoch et al. \cite{murdoch92}).  
No radial velocity monitoring is available in the literature for \siglib, 
\vcen, and \epsmus.  Therefore, for these stars, we adopted the same 
uncertainty of $\pm0.6$~\KMS\ as in \gamcru.  

On the other hand, 
the visible spectral lines in the red supergiants \alfsco\ and 
\alfher\ show time variations in the radial velocity of up to $\sim$10~\KMS\ 
due to the atmospheric motions (Smith et al. \cite{smith89}).  
We adopted these time variations as the 
uncertainty in the systemic velocity of \alfsco\ and \alfher.  
While \alfsco\ and \alfher\ both have binary companions, 
the effect of the orbital motion on the radial velocity is much smaller 
compared to the time variations due to the atmospheric motions for the 
following reason.  
The red supergiant \alfsco\ A has a B2.5V companion.  
Using the binary parameters presented in Braun et al. (\cite{braun12}), 
we estimated the orbital velocity of \alfsco\ A to be 2~\KMS, which is much 
smaller than 10~\KMS.  
The system \alfher\ includes the red supergiant $\alpha^1$~Her and 
a secondary ($\alpha^2$~Her) with a separation of 4\farcs6, 
which corresponds to 504~AU at a distance of 110~pc (from the Hipparcos 
parallax, van Leeuwen \cite{vanleeuwen07}).  
The secondary $\alpha^2$~Her itself is a spectroscopic binary 
with a G8III star ($\alpha^2$~Her A) and an A9IV-V star ($\alpha^2$~Her B) 
with a separation of 0.17~AU and a period of 51.6 days (Deutch \cite{deutch56}).  
Assuming an edge-on circular orbit around the red supergiant $\alpha^1$~Her for 
this spectroscopic binary and adopting the mass of 2.7, 2.6, and 2.0~\MSOL\ 
for $\alpha^1$~Her, $\alpha^2$~Her~A, and $\alpha^2$~Her~B, respectively, 
based on the results of Moravveji et al. (\cite{moravveji13}), 
the orbital velocity of $\alpha^1$~Her is estimated to be $\sim$2~\KMS, 
which is again much smaller than the 10~\KMS\ caused by the atmospheric 
motions. 

For \gcirs, no radial velocity measurement is available.  Therefore, 
the wavelength scale is converted only to the heliocentric frame.

\section{Results}
\label{sect_res}

\begin{figure*}
\resizebox{17.5cm}{!}{\rotatebox{0}{\includegraphics{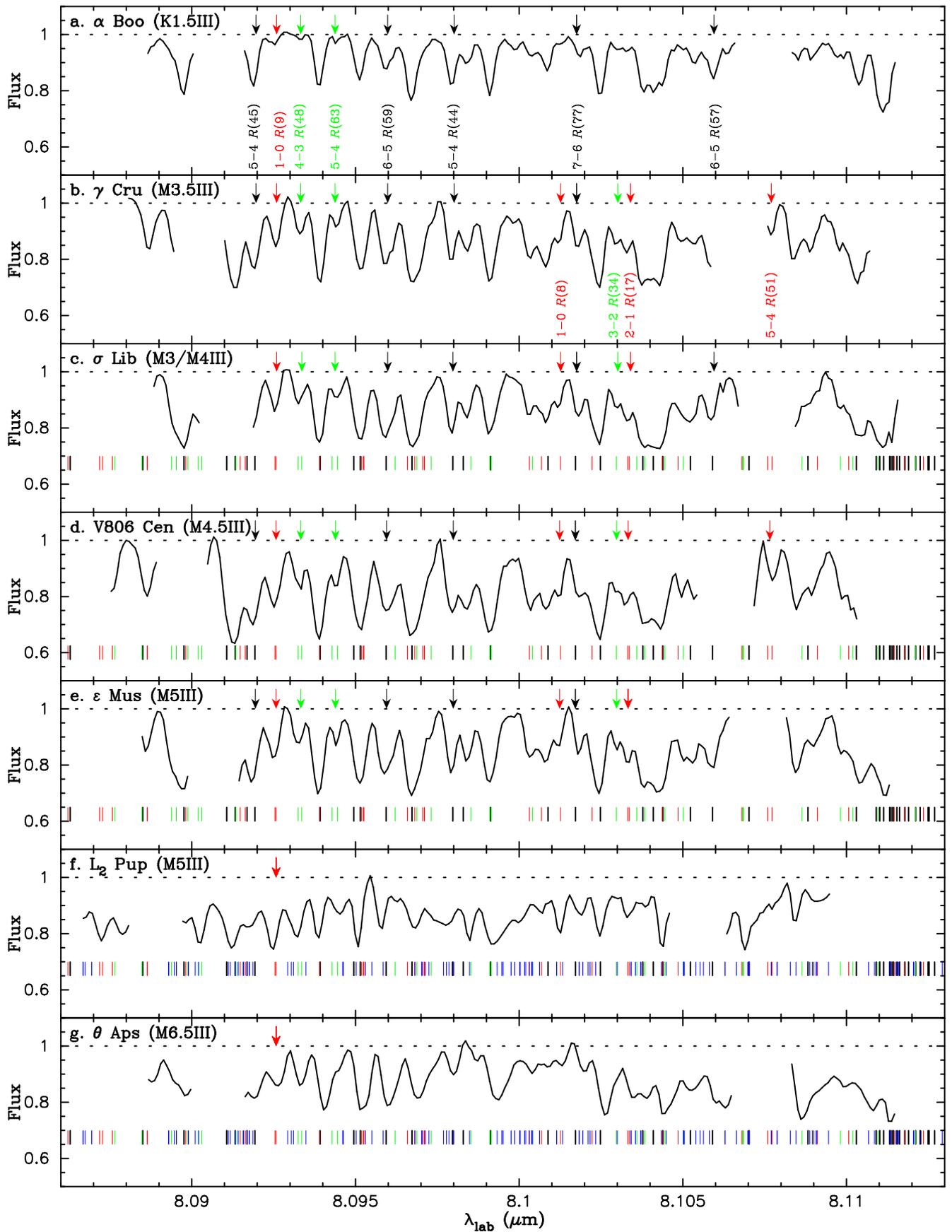}}}
\caption{
Observed VISIR spectra of the SiO fundamental lines in seven K--M giants. 
The positions of the  $^{28}$SiO, $^{29}$SiO, and $^{30}$SiO lines relatively 
free from blend with other species 
are marked by the black, red, and green arrows, respectively.  
The identification of these lines are also given (in case there are multiple 
transitions within one feature, only the main contributor is given).  
The positions of the $^{28}$SiO, $^{29}$SiO, $^{30}$SiO, and \HOH\ lines 
are shown by the black, red, green, and blue ticks, respectively 
(\HOH\ lines are plotted only for \lpup\ and \tetaps). 
The dotted lines show the local continuum.  
}
\label{obsres_giants}
\end{figure*}

\begin{figure*}
\resizebox{\hsize}{!}{\rotatebox{0}{\includegraphics{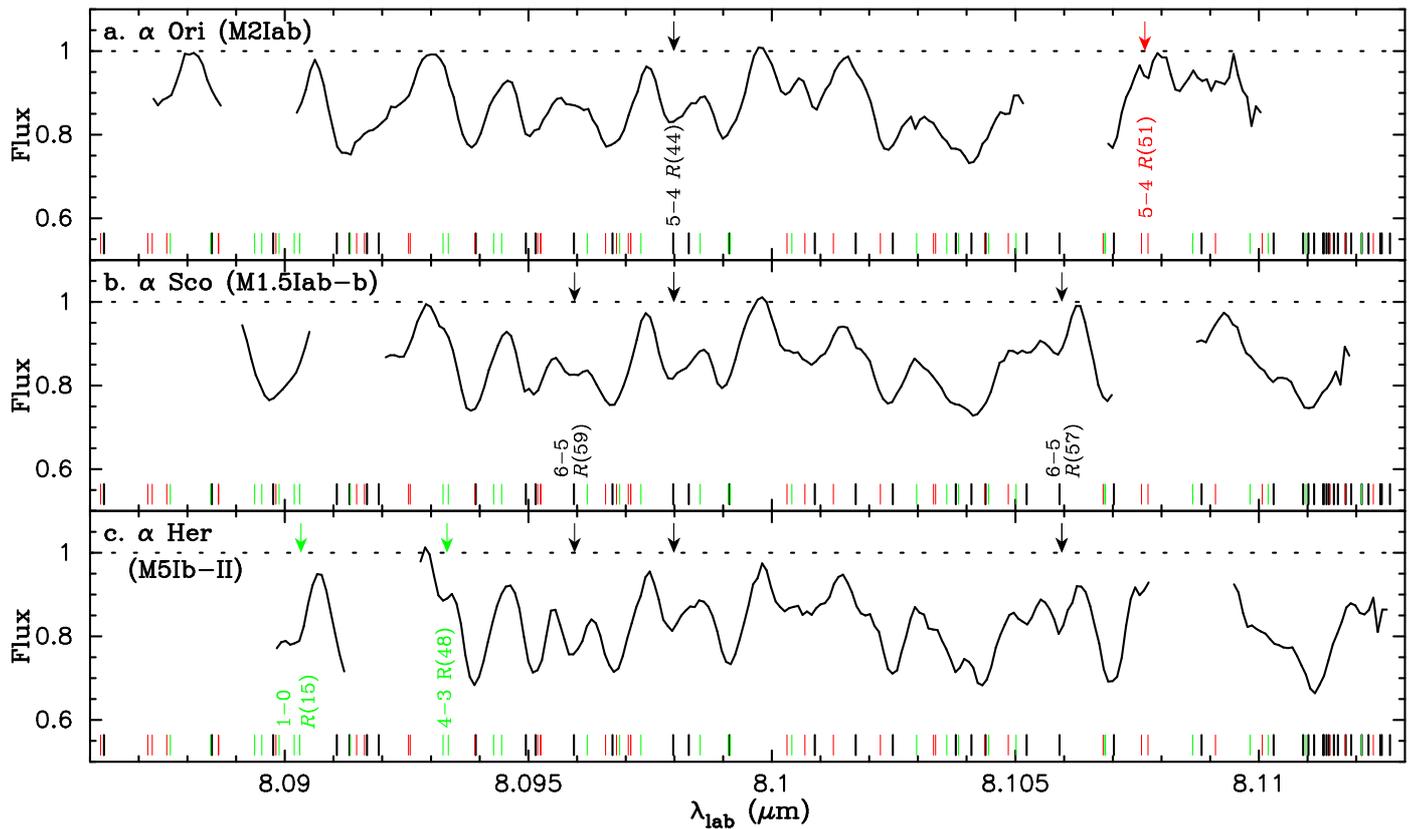}}}
\caption{
Observed VISIR spectra of the SiO fundamental lines in three optically bright 
(= not very dusty) red supergiants, shown in the same manner as in 
Fig.~\ref{obsres_giants}.  
}
\label{obsres_rsgs}
\end{figure*}

\subsection{Optically bright red giants and red supergiants}
\label{subsect_nondusty}

Figure~\ref{obsres_giants} shows the observed spectra of seven 
normal K--M giants in our sample 
with the positions of the $^{28}$Si$^{16}$O, 
$^{29}$Si$^{16}$O, and $^{30}$Si$^{16}$O lines (hereafter simply denoted 
as $^{28}$SiO, $^{29}$SiO, and $^{30}$SiO), which are taken from 
the line list recently published by Barton et al. (\cite{barton13}). 
For the spectra of \lpup\ and \tetaps, the positions of the \HOH\ 
lines taken from the line list of Barber et al. (\cite{barber06}) 
are also plotted.  
Most of the observed features are blend of 
different Si isotopes.  Several lines that are relatively 
free from the blend of other species are marked and identified 
in the figure.  
In particular, 
the $^{29}$SiO lines at 8.0925 and 8.1077~\micron, as well as 
the $^{30}$SiO lines at 8.0933 and 8.0944~\micron\ are 
nearly free from the blend of different isotope species.  
Tsuji et al. (\cite{tsuji94}) derived \SIISOTOPEF\ = 14--21 and 
\SIISOTOPES\ = 20--29 for M5--8 giants from the SiO first 
overtone lines near 4~\micron.  
We estimate the silicon isotope ratios using the observed SiO fundamental 
lines in Sect.~\ref{sect_model}.

The SiO lines can already be seen in the K1.5 giant \alfboo\ and 
become stronger in M giants (\gamcru, \siglib, \vcen, and 
\epsmus) than in \alfboo\ but without clear dependence on the spectral 
type.  
The lack of clear correlation between the strengths of the SiO fundamental 
lines and the spectral type (or effective temperature) 
is already found in the spectra with lower spectral resolutions 
(Cohen \& Davis \cite{cohen95}; Sloan \& Price \cite{sloan98}; 
Heras et al. \cite{heras02}).  
We will return to this point in Sect.~\ref{subsubsect_marcs_giants}, where 
we compare the observed data with synthetic spectra from hydrostatic 
photospheric models. 
The SiO lines become weaker in \lpup\ and \tetaps\ compared to the 
earlier M giants.  
As the D.E.C. values in Table~\ref{obslog} suggest, these 
two stars show noticeable dust emission, which makes the SiO lines appear 
weaker.  
However, as shown in Sect.~\ref{subsubsect_marcs_giants2}, 
the observed spectra of 
\lpup\ and \tetaps\ cannot be explained just by the photospheric models 
with the dust emission, 
suggesting the possible effects of the MOLsphere on 
the SiO fundamental lines.

Figure~\ref{obsres_rsgs} shows the observed spectra of the optically bright 
red supergiants \alfori, \alfsco, and \alfher.  
The SiO lines in \alfori\ and \alfsco\ are significantly 
broader compared to those observed in the K--M giants, reflecting 
the larger turbulent velocities found in these red supergiants 
(e.g., Ohnaka et al. \cite{ohnaka09} and references therein).  

The observed SiO line positions of the K--M giants and optically bright 
red supergiants agree well with 
those from the line list within the uncertainty of the wavelength 
calibration.  No P-Cyg profiles and/or blueshifts, which would 
be a signature of outflows, were detected.  
This suggests that the SiO fundamental lines form in the 
photosphere and also possibly in the MOLsphere, where the material 
is not yet systematically outflowing.

\begin{figure*}
\resizebox{\hsize}{!}{\rotatebox{0}{\includegraphics{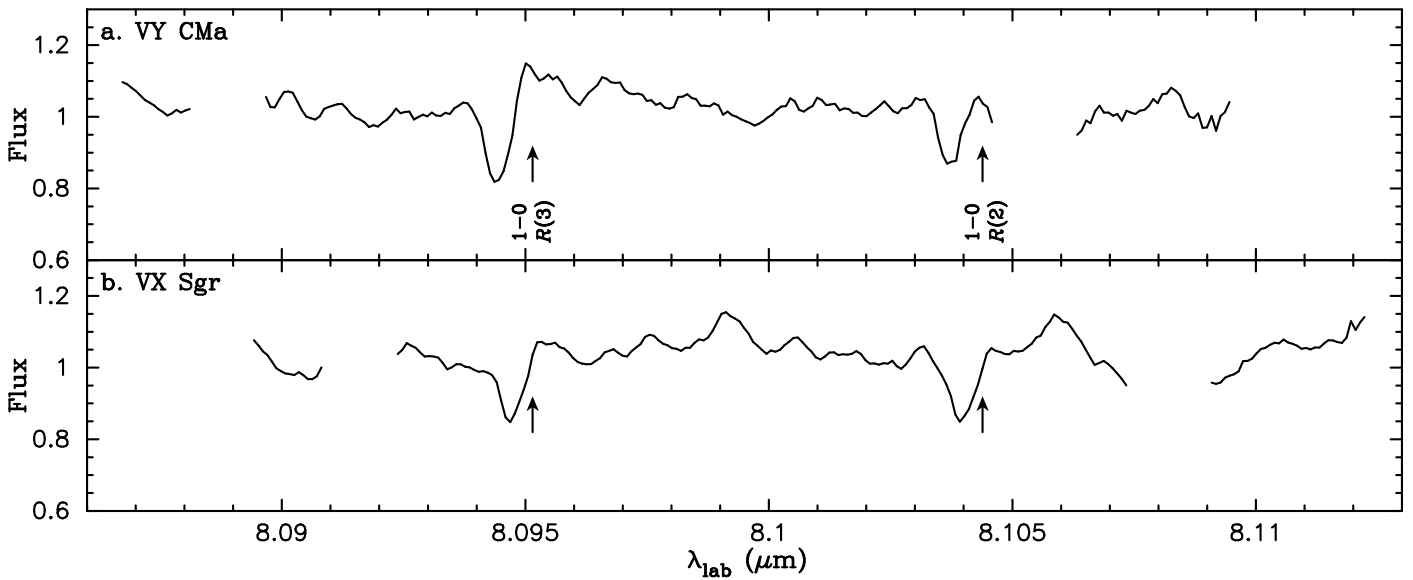}}}
\caption{
Observed VISIR spectra of the SiO fundamental lines in two dusty red
supergiants.  
The positions of the $^{28}$SiO lines with the P-Cyg profile are shown 
by the arrows.  
}
\label{obsres_dustyrsgs}
\end{figure*}

\subsection{Dusty red supergiants}
\label{subsect_dustyrsgs}

Figure~\ref{obsres_dustyrsgs} shows the observed spectra of two dusty 
red supergiants, 
\vycma\ and \vxsgr, which show much more prominent dust emission compared to 
the red giants and red supergiants discussed above.  
The observed VISIR spectra reveal the $^{28}$SiO line at 8.0952~\micron\ 
($1-0$ $R$(3)) with the P-Cyg profile.  
The mid-IR spectra of these stars are dominated by the dust emission, which 
is responsible for the continuum.  
Because the SiO lines originating in the photosphere and MOLsphere are masked 
by the prominent dust emission, the observed SiO line with the P-Cyg profile 
originates in the circumstellar envelope.  
The P-Cyg profile in the SiO line in \vycma\ is consistent with what is 
reported in Richter et al. (\cite{richter01}).  
On the other hand, the SiO line with the 
P-Cyg profile has been detected for the first time toward \vxsgr.  
Richter et al. (\cite{richter01}) report the detection of the P-Cyg profile 
in the SiO line at 8.1044~\micron\ ($1-0$ $R$(2)) as well.  
However, the line profile 
in our spectra of \vycma\ and \vxsgr\ does not indicate the P-Cyg profile 
as clearly as the SiO line at 8.0952~\micron.  This can be due to the 
lower spectral resolution ($\lambda/\Delta \lambda \approx 30000$) compared 
to the $10^5$ achieved by Richter et al. (\cite{richter01}).

\begin{figure}
\resizebox{\hsize}{!}{\rotatebox{0}{\includegraphics{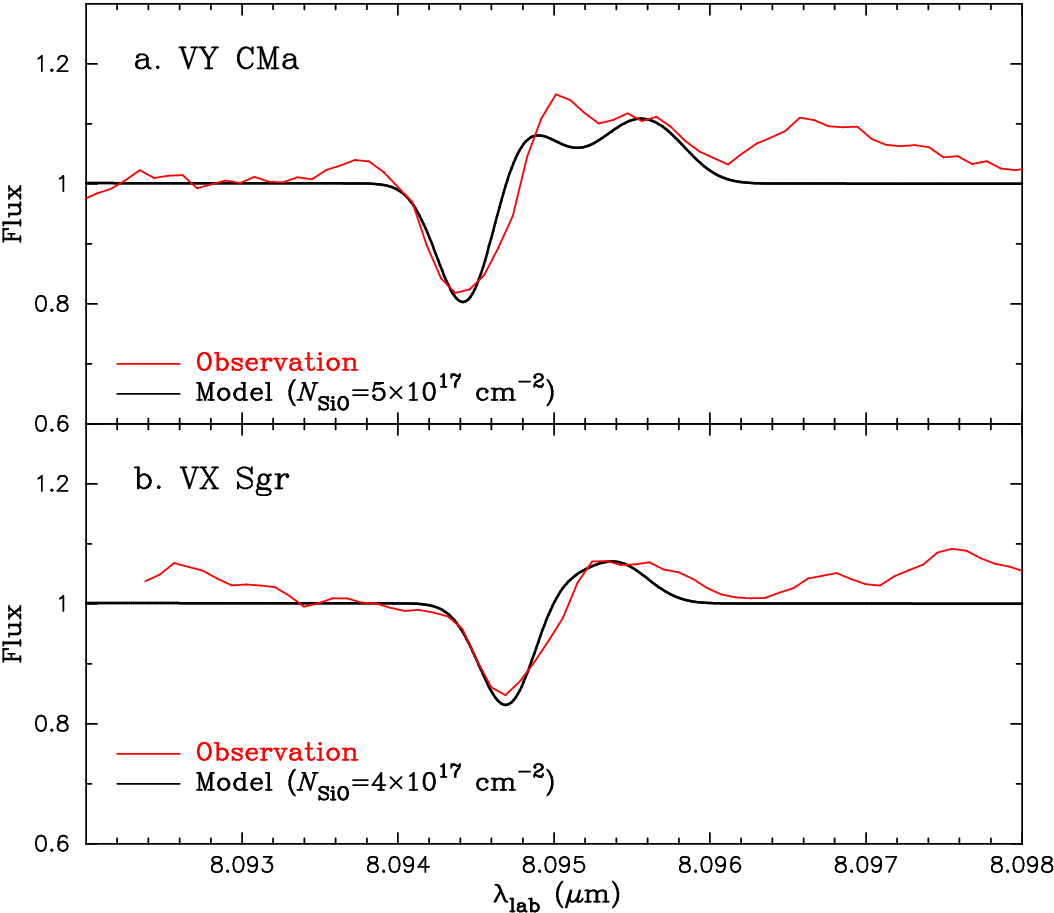}}}
\caption{
Modeling of the P-Cyg profile of the SiO line $1-0$ $R(3)$ 
observed in \vycma\ (panel {\bf a}) and \vxsgr\ (panel {\bf b}). 
In each panel, the red solid line represents the observed spectrum, 
while the black solid line represents the model spectrum. 
}
\label{model_dustyrsg}
\end{figure}

\begin{figure*}
\resizebox{\hsize}{!}{\rotatebox{0}{\includegraphics{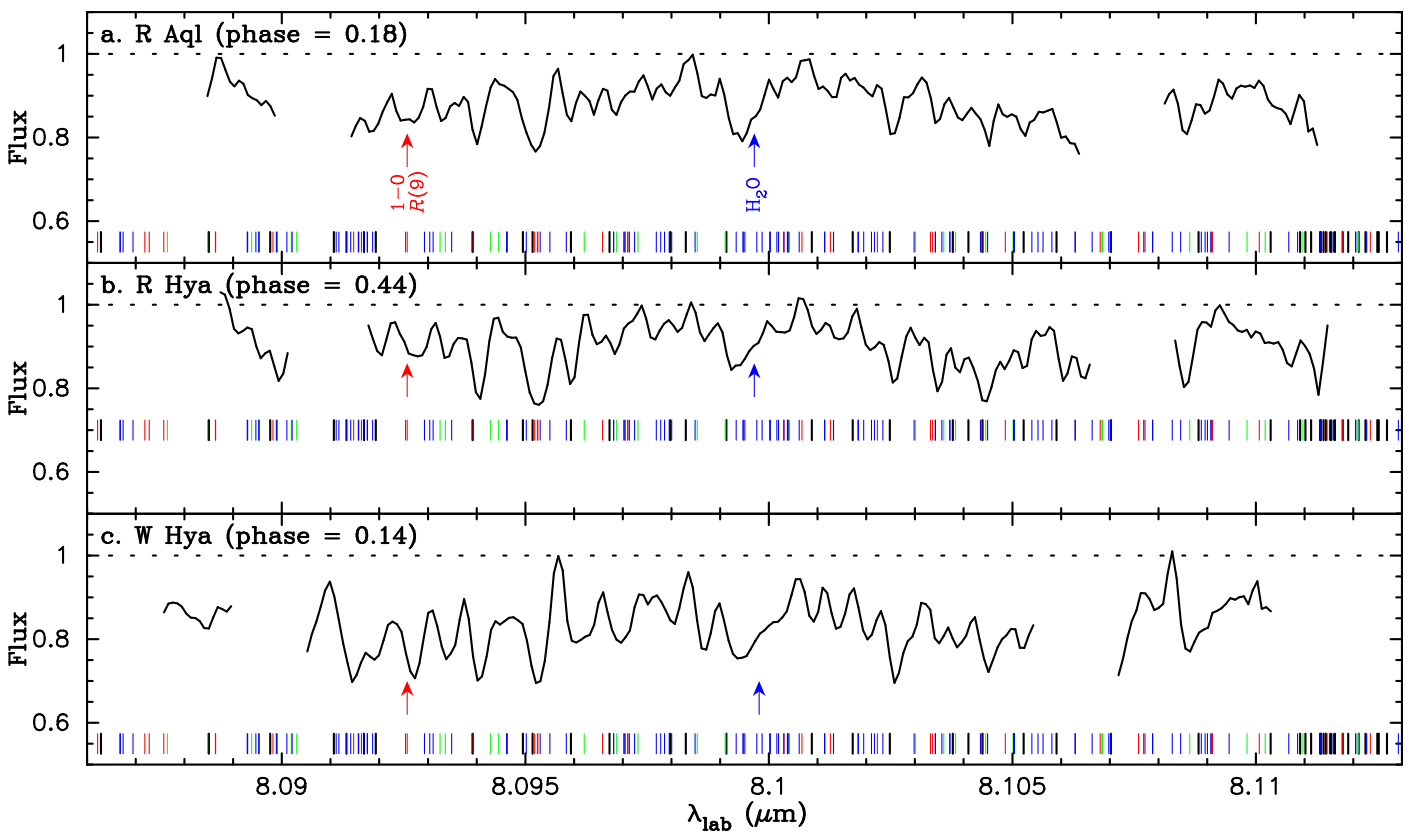}}}
\caption{
Observed VISIR spectra of the SiO fundamental lines in three Mira 
stars, shown in the same manner as in Fig.~\ref{obsres_giants}.  
}
\label{obsres_miras}
\end{figure*}

In order to estimate the outflow velocities in \vycma\ and \vxsgr, 
we computed the profile of the SiO line at 8.0952~\micron\ using 
a simple model of a spherically expanding shell with a constant velocity.  
This model consists of a dust shell, which emits as the blackbody of 1500~K 
and provides the continuum, and an expanding SiO layer with a temperature of 
500~K placed at some distance from the continuum-forming dust layer.  
The free parameters are the expansion velocity, radius, SiO column 
density, and microturbulent velocity of the SiO layer.  
The SiO line profile was computed 
assuming local thermodynamical equilibrium (LTE).  
Figure~\ref{model_dustyrsg} shows a comparison of the predicted and 
observed line profiles for \vycma\ and \vxsgr.  
The observed P-Cyg profiles of \vycma\ 
and \vxsgr\ are reasonably reproduced by an expansion velocity of 
27 and 17~\KMS, respectively.  The radius of the SiO layer in \vycma\ and 
\vxsgr\ is 2.5 and 2 times larger than the radius of the continuum-forming 
dust layer, respectively.  A microturbulent velocity of 7~\KMS\ was found 
to fit the observed profiles well.  
The SiO column density derived for \vycma\ and \vxsgr\ is $5\times10^{17}$ and 
$4\times10^{17}$~\PERSQCM, respectively.  
The column density derived for \vycma\ is in rough agreement with the 
$7\times10^{17}$~\PERSQCM\ estimated by Geballe et al. (\cite{geballe79})
from the P-Cyg profile of the SiO line at 8.309~\micron.  
However, the strength of the absorption and emission components in the P-Cyg 
profile depends on the combination of the temperature, radius, and column 
density of the SiO layer, and these parameters cannot be unambiguously 
constrained by our observations of the single SiO line alone.  
Nevertheless, the wavelength shifts of the absorption and 
emission components are primarily determined by the outflow velocity.  
Therefore, the derived outflow velocities are not significantly dependent on 
the uncertainties in the other parameters.  
Given the terminal wind velocities of 
44--48 and 22--26~\KMS\ for \vycma\ and \vxsgr, respectively 
(Kemper et al. \cite{kemper03}), this SiO line at 8.0952~\micron\ 
originates in the region, where the material has 
reached approximately 60--70\% of the terminal velocity and is being further 
accelerated.

\begin{figure*}
\resizebox{\hsize}{!}{\rotatebox{0}{\includegraphics{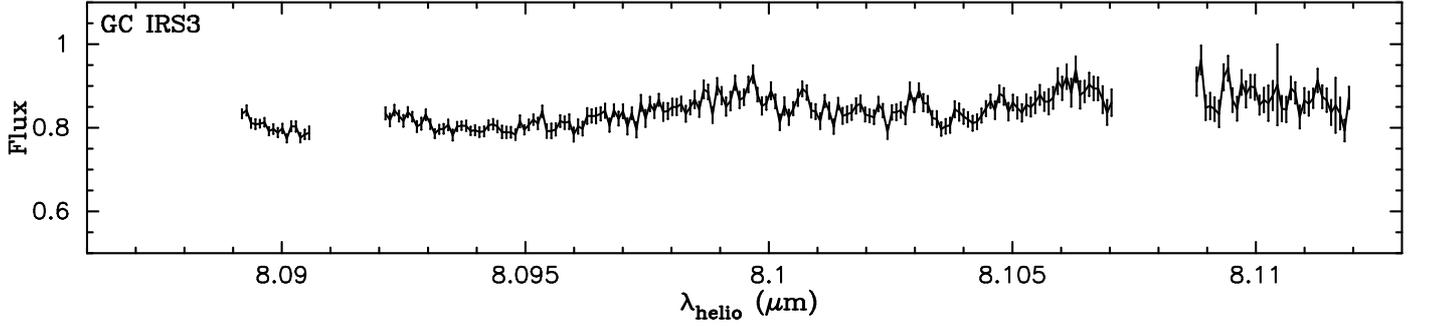}}}
\caption{
Observed VISIR spectrum of \gcirs.  
Note that the wavelength scale is heliocentric unlike in 
Figs.~\ref{obsres_giants}--\ref{obsres_miras}.  
}
\label{obsres_gcirs3}
\end{figure*}

\subsection{Mira stars}
\label{subsect_miras}

Figure~\ref{obsres_miras} shows the observed spectra of three Mira stars, 
\raql, \rhya, and \whya.  
The spectral features in R Hya and R Aql are systematically weaker compared to 
the dust-free M4--5 giants shown in Fig.~\ref{obsres_giants} 
but comparable to the M6.5 giant \tetaps.  
Given that the dust emission in \tetaps\ (D.E.C. = 0.29) is similar 
to \raql\ and \rhya\ (D.E.C. = 0.36 and 0.24, respectively), 
the contribution of the dust emission is likely (at least partially) 
responsible for making the SiO lines appear weaker 
(but note that the dust emission 
alone cannot account for the weak spectral features in \tetaps, as we 
discuss in Sect.~\ref{subsubsect_marcs_giants2}).  
This can also explain why the SiO lines in 
\whya, whose D.E.C. value (0.10) is lower than in \raql\ and \rhya, are 
more pronounced.  

As the figure shows, 
there are a number of \HOH\ lines in the observed spectral window, but 
many of them are blended with the SiO lines.  
The only \HOH\ feature mostly free from the blend with the SiO lines is 
the broad absorption between 8.0995 and 8.1~\micron.  
The only SiO line relatively free from the blend of \HOH\ lines is 
the $^{29}$SiO line at 8.0925~\micron, as marked in the figure.  
This $^{29}$SiO line in \raql\ and \rhya\ is broader 
compared to the normal K--M giants shown in Fig.~\ref{obsres_giants}. 
In the case of \rhya, the line is redshifted by 7.8~\KMS, 
although the \HOH\ lines at 8.093~\micron\ may also be partially responsible 
for the apparent wavelength shift.  
The same line in \whya\ is narrower than in \raql\ and \rhya, but is 
redshifted by 5.4~\KMS.  
While it cannot be entirely excluded that unidentified blend is responsible 
for the broadening of this $^{29}$SiO line, it may also be due to the 
line doubling caused by different radial velocities at different layers.  
For example, Hinkle \& Barnes (\cite{hinkle79}) as well as Hinkle et al. 
(\cite{hinkle82}) revealed line doubling in the near-IR \HOH\ and CO lines in 
Mira stars.  They also report time variations in the radial velocity of these 
lines with an amplitude of 10--20~\KMS.  
Therefore, the SiO fundamental lines provide a useful tool to probe the 
atmospheric dynamics.  
Monitoring observations are necessary to 
better understand the effects of the pulsation on the formation layer 
of the SiO fundamental lines.

\subsection{\gcirs}
\label{subsect_gcirs3}

The nature of \gcirs\ is not yet clear.  While the deep silicate absorption 
feature at 10~\micron\ indicates the oxygen-rich nature, Pott et al. 
(\cite{pott08}) propose that it is a dust-enshrouded carbon star 
based on the mid-IR interferometric observations.  
The observed spectrum of \gcirs\ plotted in Fig.~\ref{obsres_gcirs3} 
shows only weak features.  
We compared the spectrum of \gcirs\ with those of the 
Mira stars in our sample, but we could not identify the SiO or \HOH\ 
lines in \gcirs.  \gcirs\ does not show SiO lines in the P-Cyg profile 
unlike \vycma\ and \vxsgr, although the spectral energy distribution of 
\gcirs\ suggests that the star is dust-enshrouded.  Therefore, the 
absence of the SiO and \HOH\ lines lends support to the hypothesis 
of Pott et al. (\cite{pott08}) that 
\gcirs\ is a dust-enshrouded carbon star.

\section{Comparison with model atmospheres}
\label{sect_model}

\subsection{Determination of basic stellar parameters}
\label{subsect_param}

\subsubsection{K--M giants and optically bright red supergiants}
\label{subsubsect_param_giants}

We compared the observed SiO spectra with those predicted by photospheric 
models.  
For the normal K--M giants and optically bright red supergiants shown in 
Figs.~\ref{obsres_giants} and \ref{obsres_rsgs}, 
we used the MARCS photospheric models 
(Gustafsson et al. \cite{gustafsson08})\footnote{Available at
  http://marcs.astro.uu.se}, 
which represent plane-parallel or spherical 
non-gray hydrostatic photospheres with the molecular and atomic line opacities 
taken into account with the opacity sampling technique.  
We used spherical MARCS models for our study.  
Each MARCS model is specified by the effective temperature (\TEFF), 
surface gravity (\LOGG), stellar mass (\MSTAR), microturbulent velocity 
(\VMICRO), and chemical composition.

For two K--M giants (\alfboo\ and \siglib) and three optically bright 
red supergiants (\alfori, \alfsco, and \alfher), 
most of the basic stellar parameters, in particular \TEFF\ and \LOGG, 
are available in the literature, 
as summarized in Table~\ref{param_table}.  
For the other stars (\gamcru, \vcen, \epsmus, \lpup, and \tetaps), 
we derived \TEFF, \MSTAR, \LOGG\ in the following 
manner.  We determined \TEFF\ using the observed bolometric flux and 
the angular diameter.  
While the angular diameter measurement is available for \gamcru\ 
(Glindemann et al. \cite{glindemann01}), the angular diameter of 
the remaining stars was estimated from the relationship between the angular 
size and the $(V-K)$ color derived by van Belle (\cite{vanbelle99}). 
We used the relationship for ``normal giants and supergiants'' from 
van Belle (\cite{vanbelle99}) for \vcen, \epsmus, and \tetaps.  
The relationship for ``variable stars'' was adopted for \lpup, because this 
star shows a larger variability amplitude compared to the other stars.  
The bolometric flux of each star was computed from photometric data 
from the visible to the mid-IR ($\sim$20~\micron) 
taken from the literature 
(Johnson et al. \cite{johnson66}; Morel \& Magnenat \cite{morel78}; 
Mermilliod \cite{mermilliod91}; Ducati \cite{ducati02}; 
NOMAD Catalog, Zacharias et al. \cite{zacharias05}; 
2MASS, Skrutskie et al. \cite{skrutskie06}; 
WISE All-Sky Data Release, Cutri et al. \cite{cutri12}).  The photometric 
data were de-reddened using the interstellar extinction $A_{V}$  
estimated with the method of Arenou et al. (\cite{arenou92}) and 
the wavelength dependence of the interstellar extinction from 
Savage \& Mathis (\cite{savage79}).  
We confirmed that the \TEFF\ 
derived for \siglib\ in this manner agrees with the value determined with 
the infrared flux method (Tsuji \cite{tsuji08}), lending support to 
the reliability of our method of the \TEFF\ determination.  

The error in the derived \TEFF\ results from the uncertainty in the 
estimated angular diameter and the bolometric flux.  
Van Belle (\cite{vanbelle99}) mentions that the accuracy in the angular 
diameter derived from the $(V-K)$ color is approximately 10\%.  
The uncertainty in the bolometric flux results from the fact that 
our stars 
are semiregular or irregular variables with a 
variability amplitude of $\Delta V \la 1.5$~mag, and the photometric data 
in different 
bands were not contemporaneously taken.  We assumed an uncertainty of 13\% 
in the bolometric flux, adopting the value derived for the red supergiant 
Betelgeuse, which shows a similar variability amplitude 
(Ohnaka et al. \cite{ohnaka13alfsco}).  

We note that \lpup\ started a dimming event around 1994, which is likely 
caused by episodic dust formation (Bedding et al. \cite{bedding02}).  
The dimming event was still ongoing at the time of our VISIR 
observation.  
However, Bedding et al. (\cite{bedding02}) conclude that the 
effective temperature or the luminosity is unlikely to have changed 
compared to the pre-dimming state.  Therefore, we used the $(V-K)$ color 
of \lpup\ in the pre-dimming state to estimate \TEFF, because the 
photometric data during the dimming event are affected by the newly formed 
dust and, therefore, do not reflect the color of the star itself.

\begin{figure}
\resizebox{\hsize}{!}{\rotatebox{0}{\includegraphics{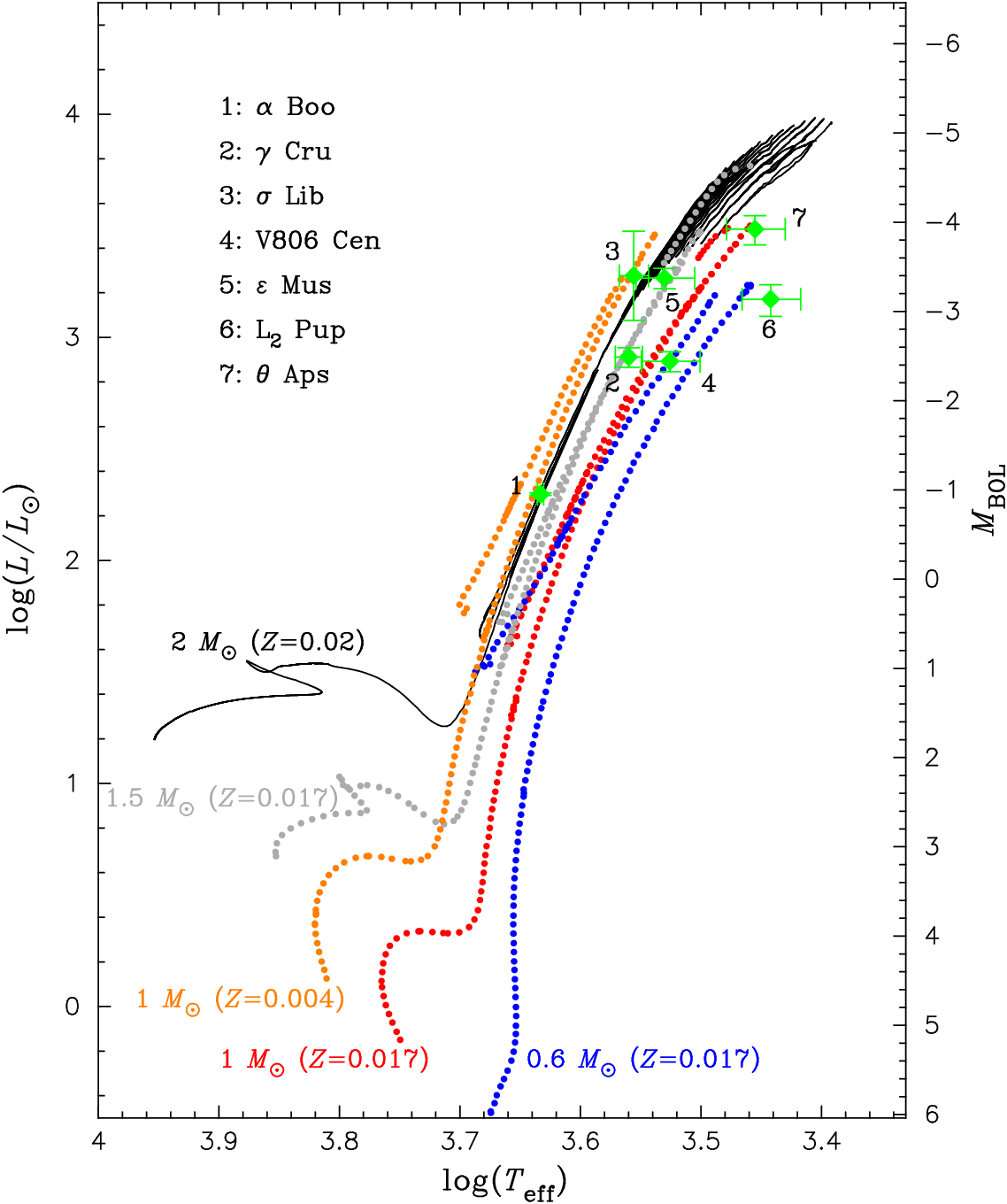}}}
\caption{
H-R diagram.  The observationally derived positions of seven K--M giants 
in our sample are plotted by the filled diamonds with error bars. 
The theoretical evolutionary track from the main sequence to the 
asymptotic giant branch for a 2~\MSOL\ star from Herwig 
(\cite{herwig05}) is shown by the solid line.  
The evolutionary tracks for 0.6, 1, and 1.5~\MSOL\ stars with the solar 
metallicity ($Z = 0.017$) and a 1~\MSOL\ star with a subsolar metallicity 
($Z=0.004$, appropriate for \alfboo) are shown by the dots.  
}
\label{hr_diagram}
\end{figure}

Because the distance to our program stars is known thanks to the 
Hipparcos parallax (van Leeuwen \cite{vanleeuwen07}), it is straightforward 
to estimate the luminosity (\LSTAR) from the de-reddened bolometric flux and 
the distance.  Using the \TEFF\ and \LSTAR\ determined in this manner, 
our program stars are plotted on the H-R diagram, as shown in 
Fig.~\ref{hr_diagram}.  Also plotted are theoretical evolutionary tracks 
for 0.6, 1, and 1.5~\MSOL\ stars taken from Bertelli et al. (\cite{bertelli08}) 
and for a 2~\MSOL\ star from 
Herwig (\cite{herwig05})\footnote{http://astrowww.phys.uvic.ca/\textasciitilde 
  fherwig}.  Comparison with these evolutionary tracks allows us to estimate 
the stellar mass.  Combining the stellar mass with the stellar radius, 
which can be computed from \TEFF\ and \LSTAR, results in the surface gravity.  

The chemical composition, in particular [Fe/H] and the CNO abundances, is 
also necessary to specify the MARCS models.  However, the analysis of 
the chemical composition is available only for two K and M giants, \alfboo\ 
and \siglib, and two red supergiants, \alfori\ and \alfher. 
The CNO abundances derived for \alfboo, \siglib, and \alfher\ 
(Decin et al. \cite{decin03}; Ramirez et al. \cite{ramirez11}; Smith et al. 
\cite{smith13}; Tsuji \cite{tsuji08}) roughly agree with the 
``moderately CN-cycled composition'' from the MARCS model grid.  
For the M giants other than \alfboo\ and \siglib, 
we assumed this chemical composition from the MARCS model grid.  
On the other hand, the CNO abundances derived for the red supergiant 
\alfori\ (Tsuji \cite{tsuji06}) are closer to the ``heavily CN-cycled 
composition'' from the MARCS grid.  For the red supergiant \alfsco, we adopt 
the same CNO abundances as \alfori, based on the similarity in \TEFF, \LOGG, 
and \MSTAR (Ohnaka et al. \cite{ohnaka13alfsco}).  
The metallicity ([Fe/H]) is determined only for \alfboo\ ([Fe/H] = $-0.5$, 
Decin et al. \cite{decin03}; Ramirez et al. \cite{ramirez11}) and 
\alfori\ ([Fe/H] = 0.1, Lambert et al. \cite{lambert84}).  
Therefore, we assumed the solar metallicity for the M giants and supergiants 
except \alfboo, for which we adopted [Fe/H] = 0.0.  

We used a microturbulent velocity of 2.5~\KMS\ for the K--M giants in our 
sample and the red supergiant \alfher\ and 5~\KMS\ for the red supergiants 
\alfori\ and \alfsco\ based on the analyses of Smith \& Lambert 
(\cite{smith90}), Tsuji (\cite{tsuji06}; \cite{tsuji08}), and 
Tsuji et al. (\cite{tsuji94}).

For each star, we selected the MARCS model with parameters closest to the ones 
derived as described above.  The derived basic stellar parameters, as well as 
the MARCS models adopted for our program stars 
are listed in Table~\ref{param_table}.

\subsubsection{Mira stars}
\label{subsubsect_param_miras}

Because the effective temperature and luminosity (and thus the radius) of 
Mira stars are expected to change with phase, it is necessary to determine the 
stellar parameters at the time of our VISIR observations to specify model 
atmospheres.  
Recently Ireland et al. (\cite{ireland11}) have presented time 
series of dynamical model atmospheres for Mira stars.  
To check whether there are models appropriate for the Mira stars that we 
observed, we estimated the luminosity of our program stars at the phase of 
our VISIR observations. 
For \raql, using the bolometric flux of 
$(351.2\pm52.7)\times 10^{-8}$~erg~cm$^{-2}$~s$^{-1}$ (Hofmann et al. 
\cite{hofmann00}) and the distance of 240~pc (Whitelock et al. 
\cite{whitelock00}), we obtain \LSTAR\ = 6290~\LSOL\ (phase = 0.18).  
For \rhya\ and \whya, we estimated the luminosity at the phase of our VISIR 
observations from the bolometric magnitude and its amplitude obtained 
by Whitelock et al. (\cite{whitelock00}).  The resulting luminosity of \rhya\ 
and \whya\ is 7200~\LSOL\ (phase = 0.44) and 8900~\LSOL\ (phase = 0.14), 
respectively.  
While some models in Ireland et al. (\cite{ireland11}) have 
approximately the same luminosities as the observationally derived ones 
at the corresponding phases, the models have either too long periods or 
too large variability amplitudes compared to \raql, \rhya, and \whya.  
The models of Ireland et al. (\cite{ireland11}) were computed for 
stellar parameters appropriate for $o$~Cet, R~Leo, and R~Cas.  
Therefore, comparison of the VISIR spectra of our Mira stars with models 
should be carried out, 
when dynamical models with parameters appropriate for our program stars become 
available.

\begin{table*}
\begin{center}
\caption {Basic stellar parameters, silicon abundance, and \SIISOTOPEF\ and
\SIISOTOPES\ ratios of the K--M giants and optically bright 
red supergiants in our sample.  The values taken from the literature are 
marked with references, while the values without notes were derived 
in the present work as described in Sect.~\ref{sect_model}. 
References: 
$a$: Smith et al. (\cite{smith13}), 
$b$: Based on the bolometric flux from Lacour et al. (\cite{lacour08}) and 
the parallax from van Leeuwen (\cite{vanleeuwen07}), 
$c$: Tsuji (\cite{tsuji08}), 
$d$: Derived from the absolute bolometric magnitude, \TEFF, and \MSTAR\ in 
Tsuji (\cite{tsuji08}), 
$e$: Ohnaka et al. (\cite{ohnaka11}), 
$f$: Derived from \MSTAR\ and the radius in Harper et al. (\cite{harper08}), 
$g$: Harper et al. (\cite{harper08}), 
$h$: Ohnaka et al. (\cite{ohnaka13alfsco}), 
$i$: Tsuji et al. (\cite{tsuji94}),
$j$: Assumed, 
$^{\dagger}$:~Heavily CN-cycled composition.  
}
\vspace*{-2mm}

\begin{tabular}{l l l l l l l l l}\hline
Name & \TEFF & \LOGG & \LSTAR & \MSTAR & MARCS model \\
     & (K)   & (cm~s$^{-2}$)& (\LSOL) & (\MSOL) & \TEFF /\LOGG /\MSTAR
/\VMICRO /[Fe/H] & \ABUNDSI & \SIISOTOPEF & \SIISOTOPES \\
\hline
\multicolumn{6}{l}{K--M giants}\\
\hline
\alfboo & $4275\pm50^{a}$ & $+1.7\pm0.1^{a}$ & $198\pm3^{b}$ &
$1.1\pm0.2^{a}$ & 4250/1.5/1.0/2.0/$-0.5$ & $7.5\pm0.1$ & $15\pm5$ & 
$25\pm10$ \\
\gamcru & $3630\pm90$ & $+0.9\pm0.1$ & $820\pm80$ & $1.5\pm0.3$       & 
3600/1.0/1.0/2.0/+0.0 & $7.5\pm0.2$ & $15\pm5$ & $25\pm10$ \\
\siglib & $3596\pm100^{c}$ & $+0.7\pm0.1^{d}$ &  $1890^{+1110}_{-700}$ $^{c}$ & 
$2.2\pm0.5^{c}$   & 3600/0.5/1.0/2.0/+0.0 & $7.5\pm0.2$ & $30\pm10$ & 
$50\pm20$ \\
\vcen  & $3360\pm190$& $+0.4\pm0.2$ & $780\pm80$ & $0.8\pm0.3$ & 
3400/0.5/1.0/2.0/+0.0 & $7.5\pm0.3$ & $15\pm5$  & $25\pm10$\\
\epsmus & $3390\pm190$& $+0.6\pm0.1$ & $1840\pm190$ & $1.5\pm0.5$     &
3400/0.5/1.0/2.0/+0.0 & $7.5\pm0.3$ & $15\pm5$ & $50\pm20$ \\
\lpup   & $2770\pm150$ & $-0.3\pm0.2$ & $1480\pm240$ & $0.5\pm0.2$   &
2800/0.0/0.5/2.0/+0.0 & --- & --- & --- \\
\tetaps & $2850\pm160$ & $-0.3\pm0.1$ & $3050\pm460$ & $1.0\pm0.3$   &
2800/0.0/1.0/2.0/+0.0 & --- & --- & --- \\

\hline
\multicolumn{6}{l}{Red supergiants}\\
\hline
\alfori & $3690\pm54^{e}$ & $-0.3\pm0.3^{f}$ & $(1.3^{+0.8}_{-0.5})\times10^5$
$^{g}$& $20\pm5^{g}$ & 3600/0.0/5.0/5.0/+0.0$^{\dagger}$ & 7.45$^{i}$ &
20$^{i}$ & 25$^{j}$ \\
\alfsco & $3660\pm120^{h}$& $-0.2\pm0.3^{h}$ &
$(7.6^{+5.3}_{-3.1})\times10^4$ $^{h}$ & $15\pm5^{h}$ &
3600/0.0/5.0/5.0/+0.0$^{\dagger}$ & 7.44$^{i}$ & 13$^{i}$ & 25$^{j}$\\
\alfher & $3293\pm100^{c}$& $+0.0\pm1.5^{d}$ & $(1.7^{+5.8}_{-1.3})\times10^4$
$^{c}$ & $5.0\pm2.0^{c}$ & 3300/0.0/5.0/2.0/+0.0 & 7.26$^{i}$ & 20$^{i}$ & 
29$^{i}$\\
\hline
\label{param_table}
\vspace*{-7mm}

\end{tabular}
\end{center}
\end{table*}

\subsection{Comparison between VISIR spectra and MARCS models}
\label{subsect_marcs}

Using the temperature and pressure stratifications of the MARCS 
models, we computed synthetic 
spectra including the SiO fundamental lines and \HOH\ lines. 
We used the SiO line list recently published by Barton et al. 
(\cite{barton13})\footnote{Available at 
http://www.exomol.com/data/molecules/SiO}, 
which is more complete than the line list of 
Langhoff \& Bauschlicher (\cite{langhoff93}) that has been widely used.  
The \HOH\ line list was taken from 
Barber et al. (\cite{barber06})\footnote{Available at 
http://www.exomol.com/data/molecules/H2O}. 
Details of the computation of the synthetic spectra are described in 
Appendix of Ohnaka (\cite{ohnaka13}).

\begin{figure*}
\sidecaption
\resizebox{12cm}{!}{\rotatebox{0}{\includegraphics{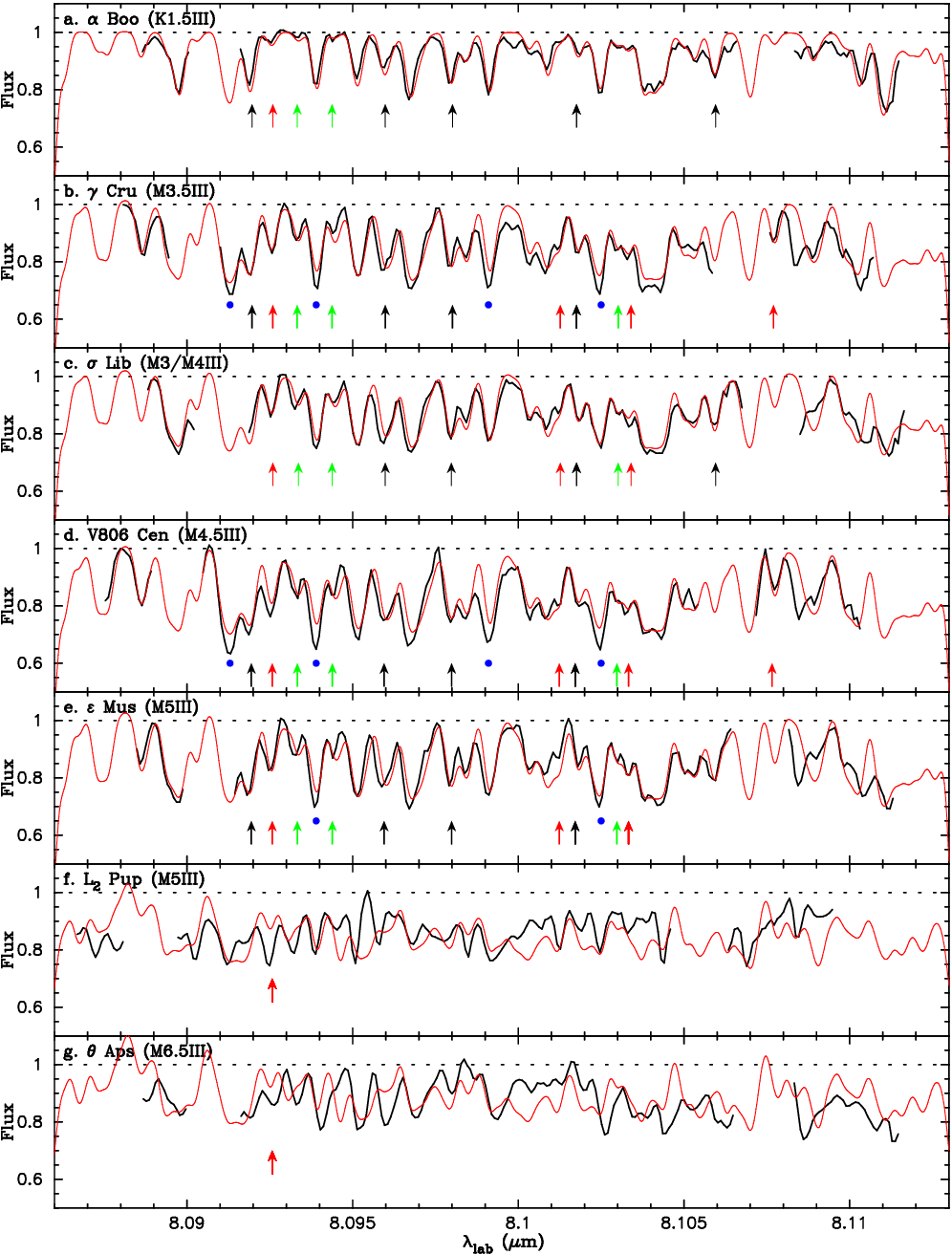}}}
\caption{
Comparison of the synthetic spectra based on the MARCS models with the 
observed VISIR spectra of seven K--M giants.  The observed and synthetic 
spectra are shown by the black and red solid lines, respectively. 
The relatively isolated $^{28}$SiO, $^{29}$SiO, and $^{30}$SiO lines are 
marked by the black, red, and green arrows, respectively.  
See Fig.~\ref{obsres_giants} for their identification. 
The $^{28}$SiO lines with absorption excess are marked by the blue dots. 
}
\label{marcs_model_giants}
\end{figure*}

\begin{figure*}
\sidecaption
\resizebox{12cm}{!}{\rotatebox{0}{\includegraphics{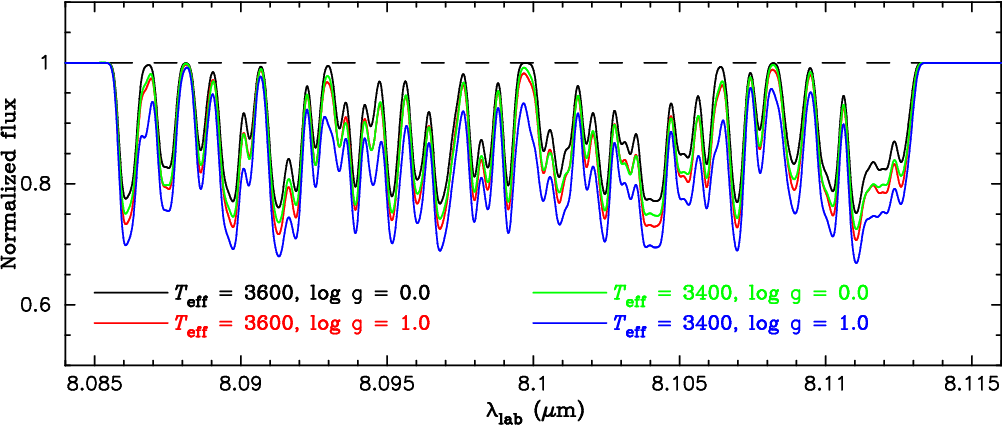}}}
\caption{
Synthetic spectra of the SiO lines 
for \TEFF\ = 3400 and 3600~K and \LOGG\ = 0.0 and 1.0.  
The dashed line represents the continuum.  We included no lines at either 
edge of the spectral range to show the continuum level.  
}
\label{synthetic_grid}
\end{figure*}

\subsubsection{K--M giants warmer than $\sim$3400~K}
\label{subsubsect_marcs_giants}

Figures~\ref{marcs_model_giants}a--e show a comparison of the VISIR spectra 
of five K--M giants warmer than $\sim$3400~K with the MARCS synthetic spectra, 
which are already convolved with a Gaussian that 
represents the instrumental resolution and the macroturbulence. 
We adopted a macroturbulent velocity of 3~\KMS\ for these K--M giants 
based on the analysis of the CO first overtone lines by 
Tsuji (\cite{tsuji86}).  

The relatively isolated $^{28}$SiO lines marked by the black arrows 
in the figure allow us to estimate the silicon abundance.  
For \alfboo, we first computed a synthetic spectrum with \ABUNDSI\ = 
7.01 (the abundance is given on the scale of \ABUNDH\ = 12), 
which corresponds to the solar value of \ABUNDSI\ = 7.51 
(Asplund et al. \cite{asplund09})\footnote{The silicon abundance 
\ABUNDSI\ = 7.51 was 
determined with a 3-D solar model atmosphere, while we used 1-D MARCS 
models.  However, Asplund et al. (\cite{asplund09}) show that the 
difference in the silicon abundance derived from the 3-D and 1-D solar 
model atmospheres is only 0.02~dex.} scaled to [Fe/H] = $-0.5$.  
However, the synthetic spectrum predicts the 
SiO lines to be too weak.  We found that the observed spectrum can be 
reproduced well by the increased silicon abundance of \ABUNDSI\ = $7.5\pm 0.1$.  
Given the error in the derived silicon abundance, 
this value roughly agrees with \ABUNDSI\ = 7.45 (errors smaller than 0.1~dex) 
derived by Peterson et al. (\cite{peterson93}) and 
\ABUNDSI\ = $7.32 \pm 0.04$ 
derived more recently by Ramirez et al. (\cite{ramirez11}) from the 
atomic Si lines in the visible.  
The observed spectra of \gamcru, \siglib, \vcen, and \epsmus\ are 
consistent with the solar silicon abundance of \ABUNDSI\ = 7.51.  
The uncertainty in the silicon abundance is 
$\pm$0.2~dex for \gamcru\ and \siglib\ and $\pm$0.3~dex for \vcen\ and 
\epsmus.  
The reason for the larger uncertainty in cooler stars is that the strength 
of the $^{28}$SiO lines observed in our spectral window become insensitive 
to the silicon abundance in cooler stars.  

Using the $^{29}$SiO lines at 8.0925, 8.1012, 8.1033, and 8.1077~\micron, 
as well as the $^{30}$SiO lines at 8.0933, 8.0944, 8.1030~\micron, 
we estimated the \SIISOTOPEF\ and \SIISOTOPES\ ratios.  
The observed $^{29}$SiO and $^{30}$SiO lines in \alfboo, \gamcru, and 
\vcen\ are reasonably reproduced by $^{28}$Si/$^{29}$Si = 15 and $^{28}$Si/$^{30}$Si 
= 25, which are the average values in M giants analyzed by Tsuji et al. 
(\cite{tsuji94}).  On the other hand, we found ($^{28}$Si/$^{29}$Si, 
$^{28}$Si/$^{30}$Si) = (30, 50) and (15, 50) for \siglib\ and \epsmus, 
respectively.  The errors in the \SIISOTOPEF\ and \SIISOTOPES\ ratios are 
$\pm$30\% and $\pm$40\%, respectively.  The relatively large errors result 
from the small number of $^{29}$SiO and $^{30}$SiO lines used 
in the analysis.  
The derived silicon abundance and isotope ratios are summarized in 
Table~\ref{param_table}.  

There are slight wavelength shifts in the observed positions of some SiO lines 
compared to the synthetic spectra (e.g., 8.094 and 8.096~\micron).  
However, given the uncertainty in the wavelength calibration of 2.6~\KMS, 
these wavelength shifts, which are 3~\KMS\ at most, cannot be definitively 
regarded as the evidence of the atmospheric motions.  

In Sect.~\ref{subsect_nondusty}, we mentioned the apparent lack of 
clear correlation between the SiO line strength and the spectral type 
among the M giants discussed 
here.  To better understand this observational result, we show 
in Fig.~\ref{synthetic_grid} the synthetic SiO 
spectra predicted from the MARCS models with \TEFF\ = 3400 and 3600~K and 
\LOGG\ = 0.0 and 1.0 (other parameters are \MSTAR\ = 1~\MSOL, 
\VMICRO\ = 2~\KMS, [Fe/H] = 0.0, and moderately CN-cycled 
composition).  The figure shows that the SiO lines become stronger with 
decreasing \TEFF\ and increasing \LOGG\ in the \TEFF\ and \LOGG\ range 
relevant to our M giants.  In our sample, the cooler stars 
tend to have lower surface gravities.  Therefore, the increase in the SiO 
line strength due to lower \TEFF\ can be canceled out by the decrease in 
the line strength due to lower \LOGG, which results in a weak dependence 
of the SiO line strength on the spectral type.

While the observed spectra of \alfboo\ and \siglib\ are explained well 
by the MARCS models, there is noticeable disagreement between the 
observed and synthetic spectra for some SiO lines in \gamcru, \vcen, 
and \epsmus.  
The $^{28}$SiO lines at 8.0939 (2--1 $R(12)$), 8.0991 (3--2 $R(21)$), 
and 8.1025~\micron\ (2--1 $R(11)$), marked by the blue dots in 
Fig.~\ref{marcs_model_giants}, are stronger than predicted by the MARCS 
models. 
Increasing the silicon abundance does not reconcile this problem, because 
it would worsen the agreement for the other $^{28}$SiO lines marked by 
the black arrows.  The lines at 8.0939 and 8.0991~\micron\ include blend of 
weaker $^{29}$SiO and $^{30}$SiO lines, respectively.  However, increasing 
the abundance of these isotopic species would worsen the agreement for 
other $^{29}$SiO and $^{30}$SiO lines marked by the red and green arrows. 
Changing the stellar parameters within their uncertainties does not 
improve the agreement either.  

The $^{28}$SiO lines with the absorption excess have low excitation potentials 
(0.16--0.30~eV), while the weaker $^{28}$SiO lines reproduced well by the 
models have higher excitation potentials of 0.77--1.40~eV.  
This is similar to what is found for the SiO first overtone lines 
in latest M giants by Tsuji et al. (\cite{tsuji94}), which is interpreted 
as the indication of the MOLsphere, as outlined in Sect.~\ref{sect_intro}.  
Therefore, the absorption excess in the SiO fundamental lines with low 
excitation potentials may also be a signature of the MOLsphere.  
However, as described in Sect.~\ref{subsect_param}, we assumed the CNO 
abundances and [Fe/H] for \gamcru, \vcen, and \epsmus, because no quantitative 
analysis of the chemical composition is available in the literature.  
Therefore, we cannot exclude the possibility that the absorption excess may be 
due to this assumed chemical composition.  
To confirm that the absorption excess is indeed a signature of the MOLsphere, 
the determination of the CNO abundances and [Fe/H] in these stars is 
necessary.

\begin{figure*}
\sidecaption
\resizebox{12cm}{!}{\rotatebox{0}{\includegraphics{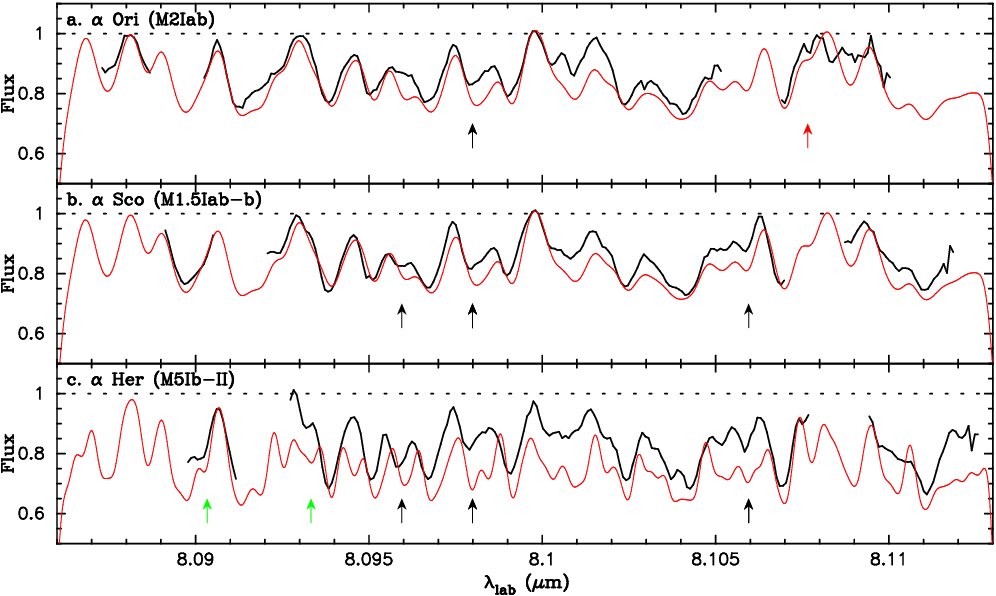}}}
\caption{
Comparison of the synthetic spectra from the MARCS models with the 
observed VISIR spectra of three optically bright red supergiants, shown 
in the same manner as in Fig.~\ref{marcs_model_giants}.  
See Fig.~\ref{obsres_rsgs} for the identification of the lines.  
}
\label{marcs_model_rsgs}
\end{figure*}

\subsubsection{Cool M giants with $\TEFF \approx 2800$~K: \lpup\ and \tetaps}
\label{subsubsect_marcs_giants2}

As Figs.~\ref{marcs_model_giants}f and \ref{marcs_model_giants}g show, 
the agreement between the observed and synthetic spectra is very poor for 
\lpup\ and \tetaps.  In these stars, which are much cooler (\TEFF\ = 
2800--2900~K) than the stars discussed above, 
the MARCS models predict the \HOH\ lines to be strong. 
Most of the features in the synthetic spectra shown in 
Figs.~\ref{marcs_model_giants}f and \ref{marcs_model_giants}g are due to 
\HOH.  
However, the shape and depth of the 
spectral features predicted by the models show little correspondence to 
the observed spectra.  

Because \lpup\ and \tetaps\ show dust emission (D.E.C. = 0.60 and 0.29, 
respectively), we tentatively added a fractional contribution of the continuum 
dust emission of 50\% in the synthetic spectra. 
Changing the fractional contribution of the dust continuum emission does not 
reconcile the disagreement, because it only makes all 
spectral features appear weaker or stronger and cannot alter the relative 
strengths of different spectral features.  
On the other hand, 
if there is emission from the extended MOLsphere, 
it can fill in the absorption and make the resultant spectrum 
(i.e., integrated over the entire stellar image in the sky) appear very 
differently from the photospheric spectrum.  
Therefore, the disagreement between the observed and photospheric 
model spectra in \lpup\ and \tetaps\ may be, again, a signature 
of the MOLsphere.  

In Figs.~\ref{marcs_model_giants}f and \ref{marcs_model_giants}g, 
the normalized flux of the synthetic spectra 
is higher than 1 at some wavelengths that are masked by the strong telluric 
lines (e.g., 8.088~\micron).  This is not emission but simply because 
we attempted to obtain the best match to the VISIR data at the wavelengths 
not severely affected by the telluric lines.  
In other words, the wavelengths of the strong telluric lines 
were excluded from the fitting.

\subsubsection{Optically bright red supergiants}
\label{subsubsect_marcs_rsgs}

Figure~\ref{marcs_model_rsgs} shows a comparison between the observed 
and synthetic spectra for the three optically bright red supergiants.  
For \alfori, \alfsco, and \alfher, we adopted 
(\ABUNDSI, \SIISOTOPEF, \SIISOTOPES) = (7.45, 20, 25), (7.44, 13, 25), 
and (7.26, 20, 29), respectively, 
based on the results from the SiO first overtone lines by Tsuji et al. 
(\cite{tsuji94}). 
Because the \SIISOTOPES\ ratio could not be determined for \alfori\ and 
\alfsco\ in Tsuji et al. (\cite{tsuji94}), we assumed it to be the same 
as the average value among M giants.  
Note that these silicon abundances were already 
corrected for the difference in the $\varg$$f$-values used in 
Tsuji et al. (\cite{tsuji94}) and this work.  
We adopted a macroturbulent velocity of 10~\KMS\ for \alfori\ 
(Tsuji \cite{tsuji06}) and assumed the same value for \alfsco, 
while we used a macroturbulent velocity of 3~\KMS\ for \alfher\ 
(Tsuji \cite{tsuji86}).  
The figure shows that the VISIR spectra of \alfori\ and \alfsco\ are 
fairly reproduced by the MARCS models, although the agreement 
is not as good as in the warm K--M giants discussed in 
Sect.~\ref{subsubsect_marcs_giants}. 
The agreement is much worse for \alfher.  The model predicts many 
features to be too strong.  We computed synthetic spectra using models 
with slightly different parameters but could not improve the fit.  
Given that \alfher\ shows little dust emission, 
the disagreement cannot be solved by adding the dust emission either.  
As in the case of \lpup\ and \tetaps\ discussed above, 
this disagreement may be attributed to the emission from the extended 
MOLsphere.  

The fair agreement between the observed and model spectra for \alfori\ is 
a little surprising, because Tsuji (\cite{tsuji06}) demonstrates that 
the SiO absorption bands observed longward of 7.5~\micron\ with a spectral 
resolution of 1600 are much weaker than predicted by the photospheric model, 
suggesting the contribution of \HOH\ and possibly also SiO emission from the 
extended MOLsphere.  
However, the signature of the MOLsphere can be masked in spatially 
unresolved spectra, because the additional absorption due to the MOLsphere 
can be filled in by the extended emission from the MOLsphere itself.  
This can make the resultant, spatially unresolved spectrum appear nearly 
unchanged.  
For example, Ohnaka et al. (\cite{ohnaka12}) and Ohnaka (\cite{ohnaka13}) 
demonstrate that although the spatially unresolved spectra of the 
2.3~\micron\ CO first overtone lines in K and M giants can be reproduced 
well by the MARCS photospheric models alone and show very little signature 
of the MOLsphere, spatially resolved spectro-interferometric 
observations clearly reveal the presence of the extended MOLsphere.  
This means that constraining the parameters of the MOLsphere from spatially 
unresolved spectra alone is not necessarily straightforward.  

Therefore, 
to probe the properties of the MOLsphere from the SiO fundamental lines, 
it would be useful to combine the present VISIR spectra with interferometric 
observations. 
The mid-IR interferometric instrument MIDI at the ESO's Very Large Telescope 
Interferometer (VLTI), as well as the next generation VLTI instrument MATISSE, 
can spatially resolve our program stars from 8 to 13~\micron.  
Although the spectral resolution is much lower ($\sim$200 at most with 
MIDI), combination of the spatially unresolved, high spectral resolution 
VISIR data and spatially resolved data with a lower spectral resolution taken 
with MIDI or MATISSE would enable us to constrain the parameters of the 
MOLsphere more reliably.  We plan to present such an analysis for \alfsco\ 
and \alfher, for which we have obtained MIDI data, in a forthcoming paper.

\section{Concluding remarks}
\label{sect_concl}

We have obtained high spectral resolution spectra of the 8.1~\micron\ 
SiO fundamental lines in 16 cool evolved stars, 
consisting of seven normal (= non-Mira) K and M giants, five red supergiants, 
three Mira stars, and the enigmatic object \gcirs\ toward the Galactic 
center. 
The SiO lines detected in the K--M giants, 
as well as in the optically bright red supergiants, do not show a signature of 
systematic outflows.  
On the other hand, 
we detected SiO lines with the P-Cyg profile in the dusty red supergiants 
\vycma\ and \vxsgr, with the latter object being a new detection.  
A simple modeling of the observed P-Cyg profiles suggests an outflow velocity 
of 27 and 17~\KMS\ in \vycma\ and \vxsgr, respectively.  
We also detected a broadening and/or redshift in an $^{29}$SiO line in Mira 
stars.  
In \gcirs, no SiO or \HOH\ lines can be identified, 
which is consistent with the previous suggestion that this 
object is a dust-enshrouded carbon star, despite the presence of 
the 10~\micron\ silicate feature. 

The observed spectra of K--M giants warmer than $\sim$3400~K 
are reasonably reproduced by the MARCS photospheric models.  
On the other hand, the MARCS models cannot explain the observed spectra of 
the cooler (\TEFF\ $\approx 2800$~K) M giants at all, even if the 
dust emission is taken into account. 
Similarly, while the observed spectra of the red supergiants \alfori\ and 
\alfsco\ (\TEFF\ $\approx 3700$~K) can be fairly reproduced by the MARCS 
models, the observed spectrum of the cooler red supergiant \alfher\ 
(\TEFF\ $\approx 3300$~K) 
shows noticeable disagreement with the MARCS model.  
This disagreement in the cooler M giants and red supergiant may be a 
signature of the MOLsphere.  
Combination of VISIR high resolution spectroscopy and mid-IR 
spectro-interferometry would be necessary to probe the contribution of the 
MOLsphere in the SiO fundamental lines.  
Given that silicon plays an 
important role in dust formation in oxygen-rich stars, studying 
the physical properties of the SiO gas in the MOLsphere is 
indispensable for understanding the dust formation in oxygen-rich stars.

\begin{acknowledgement}
We thank the ESO VLT team for supporting our VISIR observations.  
This research has made use of the SIMBAD database,
operated at CDS, Strasbourg, France.  
We acknowledge with thanks the variable star observations from the AAVSO 
International Database contributed by observers worldwide and used in this 
research.  
This publication makes use of data products from the Wide-field
Infrared Survey Explorer, which is a joint project of the University of
California, Los Angeles, and the Jet Propulsion Laboratory/California
Institute of Technology, funded by the National Aeronautics and Space
Administration.
This publication also makes use of data products from the Two Micron All Sky
Survey, which is a joint project of the University of Massachusetts and the
Infrared Processing and Analysis Center/California Institute of Technology,
funded by the National Aeronautics and Space Administration and the National
Science Foundation.
\end{acknowledgement}

\end{document}